\documentclass[epj]{webofc}
\usepackage[varg]{txfonts}   
\usepackage{hyperref}
\usepackage{textcomp}

\renewcommand*{\eqref}[1]{Eq.~(\ref{eq:#1})}

\newcommand*{\figref}[1]{Fig.~(\ref{fig:#1})}
\newcommand*{\figlab}[1]{\label{fig:#1}}

\wocname{\includegraphics[width=0.25cm,clip]{logoARENA} ARENA2018}
\wocname{ARENA2018}
\woctitle{ARENA2018}
\woctitle{\includegraphics[width=0.25cm,clip]{logoARENA} ARENA2018}
\begin{document}
\title{Reconstruction of air-shower measurements with AERA in the presence of pulsed radio-frequency interference}
%
%

\author{
\firstname{Tim} 
\lastname{Huege}\inst{1,4}\thanks{\email{tim.huege@kit.edu}},
\firstname{Christoph B.} \lastname{Welling}\inst{2}\ 
for the Pierre Auger Collaboration\inst{3}\fnsep\thanks{Full author 
list at \url{http://www.auger.org/archive/authors_2018_06.html}}
}


\institute{Karlsruhe Institute of Technology, Institute for Nuclear 
Physics, Karlsruhe, Germany
\and
RWTH Aachen University, Aachen, Germany
\and 
Observatorio Pierre Auger, Av. San Mart\'in Norte 304, 5613 Malarg\"ue, Argentina
\and       also at: Vrije Universiteit Brussel, Brussels, Belgium
}

\abstract{%
The Auger Engineering Radio Array (AERA) is situated in the Argentinian 
Pampa Amarilla, a location far away from large human settlements. 
Nevertheless, a strong background of pulsed radio-frequency 
interference (RFI) exists on site, which not only makes radio self-triggering
challenging but also poses a problem for an efficient and pure reconstruction
of air-shower measurements. We present how our standard event reconstruction
exploits several strategies to identify and suppress pulsed noise, 
and quantify the efficiency and purity of our algorithms.
These strategies can be employed by any experiment taking radio data in the
presence of pulsed RFI.
}
\maketitle
\section{Introduction} \label{intro}

AERA consists of 153 radio-detector stations covering a total area of 
17~km$^2$ in the Pampa Amarilla near Malargüe, Argentina. Although situated in a 
rural area, a strong background of pulsed RFI 
is present at the site. Investigations of its origin were not entirely 
conclusive, but it is clear that the pulses arise from multiple 
sources: power lines, faulty transformers in a nearby village, 
communication towers, oil rigs, and others. The average rate of
background pulses above a signal-to-noise threshold of 10 amounts to 15~kHz,
and the pulse characteristics are generally  similar to signals emitted by extensive air showers,
making, in particular, the self-triggering on radio signals very 
challenging.

While the exploitation of external triggering from other detector components of the
Pierre Auger Observatory provides an efficient and pure trigger, the 
pulsed RFI also poses a challenge for successful event 
reconstruction. In the following, we first describe how we exploit the hybrid 
nature of our measurements in the event reconstruction to mitigate the 
adverse effects of pulsed RFI. Afterwards, we 
describe our strategies to identify and reject pulsed RFI disturbing 
the measurements in individual AERA 
antenna stations, which improves both the efficiency and purity of our 
reconstruction. 

\section{Hybrid reconstruction strategy}

The Pierre Auger Observatory is a hybrid detector measuring individual 
air showers with complementary detection techniques 
\cite{AugerNIM2014}. The AERA events triggered by the Auger surface 
detector (SD) also have an associated SD measurement. Reconstruction of 
the SD data provides the event geometry, i.e., the orientation and impact 
point of the air-shower axis. Using this information early on in the 
radio reconstruction, we calculate for each AERA antenna station the 
expected time of arrival of the radio pulse. Uncertainties in the 
SD-reconstructed event geometry are propagated to set the width of the
``signal-search window'', which typically amounts to $\sim 1000$~ns.

This strategy is vastly superior to searching for radio pulses in the 
whole recorded radio trace of length 10~\textmu s, which for 15~kHz of RFI 
pulses has a probability of 15\% to harbor an RFI pulse in each 
individual antenna station. As typically 
more than 100 antenna stations are read out on an SD trigger, this 
on average would yield $\sim 15$ stations with RFI pulses, while cosmic-ray pulses are 
typically only present in 3-5 AERA stations \cite{AERAEnergyPRD}.

\section{Station-level suppression of RFI pulses}

With a signal search window of $\sim 1000$~ns, the chance probability of 
picking up an RFI pulse shrinks to $\sim 1.5$\%. For more than 100 stations 
read out, typically one or two RFI pulses remain in the data 
used for reconstruction. These pulses deteriorate the reconstruction 
efficiency and purity. We have thus developed several strategies to 
identify and suppress these RFI pulses.

The strategies have been evaluated by performing a standard event 
reconstruction on CoREAS simulations \cite{HuegeARENA2012a} which have undergone a 
complete detector simulation \cite{AbreuAgliettaAhn2011} and have been 
added to measured noise. An RFI pulse 
correctly identified as RFI is denoted a ``correct rejection'' in the 
following. Correspondingly, a ``false rejection'' characterizes a 
cosmic-ray pulse identified as RFI. For each strategy discussed in the 
following, we evaluate the fraction of correct versus false rejections 
individually before combining all algorithms and evaluating the overall 
performance.

\begin{figure*}[tb]
\centering
\includegraphics[width=0.71\textwidth,clip]{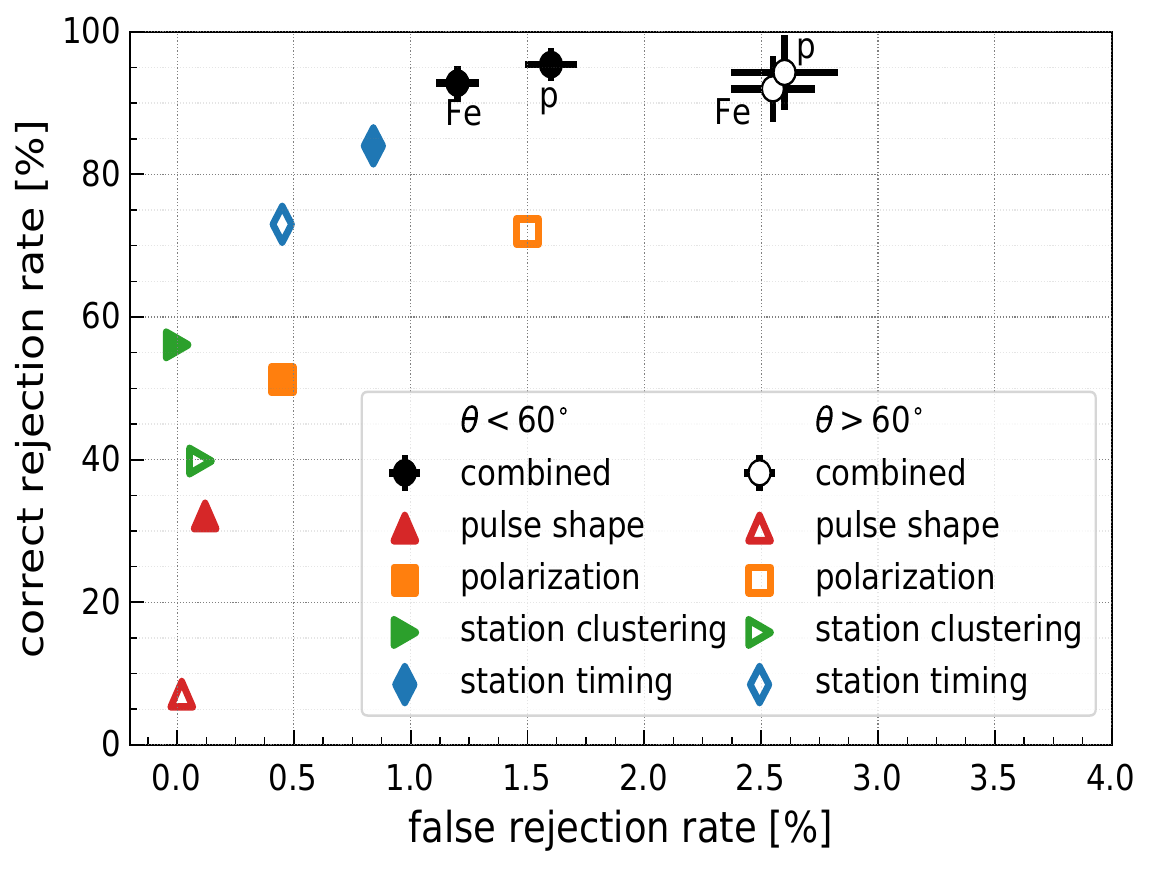}
\caption{Performance of the individual rejection strategies for air 
showers with zenith angles below (filled symbols) and above 
$60^{\circ}$ (open symbols). The black points illustrate the 
performance of all strategies combined in the same reconstruction.}
\figlab{Performance}       
\vspace{-5mm}
\end{figure*}

{\bf Pulse shape:} Radio pulses from air showers measured in the 30 to 80~MHz band are 
typically bandwidth-limited with a characteristic width and shape, oscillating with a
frequency of 55~MHz. 
We define two thresholds: a fraction $a$ of the peak amplitude in the 
signal-search window, and
a multiple $b$ of the trace RMS in a noise window before the arrival of the 
pulse. The algorithm counts the upward crossings $N$ of these two thresholds 
in a given time window $\Delta t$ centered on the pulse. Pulses with 
more crossings than a threshold $N_{\mathrm{max}}$ are too long and thus rejected as RFI.
After optimization of the parameters $a, b, \Delta t$ and 
$N_{\mathrm{max}}$, the performance is shown in the red triangles in 
\figref{Performance}.

{\bf Signal polarization:} Radio pulses from air showers have specific polarization 
characteristics arising from the superposition of geomagnetic and 
charge-excess radiation (see, e.g., \cite{HuegePLREP}). Using the event 
geometry provided by the SD reconstruction (including its uncertainties),
the actual pulse polarization at each antenna station is compared with
the expected polarization. If the deviation is larger than a defined 
threshold, the pulse is considered RFI and is rejected. The performance 
is demonstrated by the yellow squares in \figref{Performance}.

{\bf Station clustering:} The radio-emission footprint covers a contiguous area. 
Consequently, isolated antenna stations with a detected pulse have 
likely measured RFI and are rejected from the reconstruction. 
The performance of this algorithm is illustrated by the green triangles 
in \figref{Performance}.

{\bf Station timing:} The arrival times of radio pulses from an air 
shower form a wavefront. We perform a bottom-up reconstruction which first fits 
the radio-pulse arrival times of the three antenna stations nearest the 
SD core with a plane wave and then subsequently adds antenna stations further out to 
the fit. If adding a particular antenna station decreases the $\chi^2$ 
probability of the fit below 5\%, that station is rejected -- its pulse 
arrival time does not match that of the other pulses. The performance 
is illustrated by the blue diamonds in \figref{Performance}.

{\bf Overall performance:} We now combine all of these strategies in 
the reconstruction and apply them to a data set of simulations with both 
proton and iron primaries. The overall performance is shown in the 
black points in \figref{Performance}. Approximately 90\% of the RFI 
pulses are rejected while only $\sim 1.5$\% of the radio pulses from 
vertical and $\sim 2.5$\% of the radio pulses from inclined air showers 
are rejected. There are only small differences in the performance for 
proton- and iron-induced air showers, i.e., the RFI-rejection 
algorithms do not introduce a significant selection bias.

{\bf Validation with measured data:} We have validated the performance 
of the combination of these algorithms also with measured data. For 
each measured air shower, we determine the probability for successful reconstruction 
using the SD-reconstructed geometry and energy as input for an analytical model \cite{Glaser:2018byo}.
We find that for events for which 
no successful detection is expected, the fraction of reconstructed 
versus triggered events drops by a factor of $\sim 6$ when applying the 
RFI rejection. This illustrates 
that in particular the purity of our reconstruction is significantly 
increased.

\section{Conclusions}

We have presented how we achieve efficient and pure reconstruction of 
radio data taken in the presence of strong pulsed RFI by exploiting 
hybrid measurements. The strategies 
employed here can also be of use to other radio-detection projects measuring in noisy 
environments.



%
%
%
%
%

\end{document}